\documentclass[preprint,12pt]{elsarticle}
\usepackage{times} 
\usepackage{amssymb}
\usepackage{amsmath}
\usepackage[utf8]{inputenc}
\usepackage{graphicx} 
\biboptions{sort&compress,semicolon}

\journal{JMMM}

\begin{document}
	
\begin{frontmatter}
\title{Approximate explicit formulas for Stoner–Wohlfarth hysteresis loops}
	
\author[1]{Vladimir P. Savin\corref{cor1}}
\ead{savinvp99@gmail.com}

\author[1,2,3]{Yury A. Koksharov}

\cortext[cor1]{Corresponding author}

\affiliation[1]{organization={Faculty of Physics, Lomonosov Moscow State University},
	postcode={119991},
	city={Moscow},
	country={Russia}}
\affiliation[2]{organization={Kotelnikov Institute of Radio-Engineering and Electronics, Russian Academy of Sciences},
	city={Moscow},
	postcode={125009},
	country={Russia}}
\affiliation[3]{organization={Faculty of Physics and Mathematics, Shenzhen MSU-BIT},
	city={Shenzhen},
	postcode={518179},
	country={China}}
	
\begin{abstract}
Approximate explicit formulas for the hysteresis loops in the Stoner–Wohlfarth model are derived. We consider the hysteresis loops both for a single particle with a fixed easy-axis direction and for an ensemble of particles with randomly oriented anisotropy axes. The physical assumption used to derive the formulas is that the particle magnetic moment lies in the vicinity of the easy axis or the external field direction, at low and high fields, respectively. Surprisingly, the low-field formula is approximately valid even near the Stoner–Wohlfarth astroid, where the reduced magnetic field $h_0$ is not very small. The general piecewise formula is obtained by an appropriate matching of the functions defined on different intervals of the magnetic field, which are chosen to maximize the formula accuracy. For the averaged hysteresis loop, the maximal, but reasonably small, deviation of our formula from numerically calculated magnetization occurs at $h_0 \approx 0.5$, which corresponds to the sharp change in magnetization slope. 
\end{abstract}

\begin{graphicalabstract}
	\includegraphics[width=\linewidth, keepaspectratio]{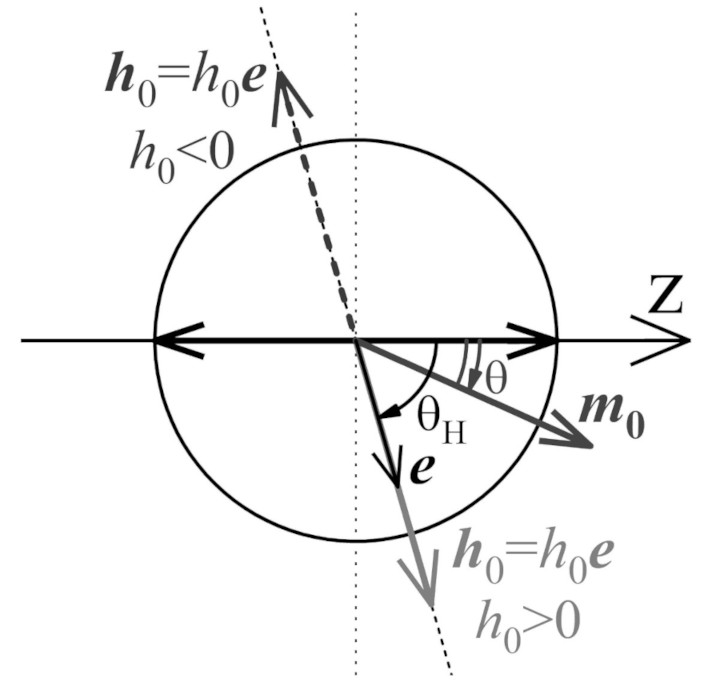}
\end{graphicalabstract}

\begin{highlights}
	\item  Explicit approximate formulas for Stoner–Wohlfarth hysteresis loops are derived, assuming that at low fields the magnetic moment aligns with the easy axis, while at high fields it aligns with the external field.
	
	\item The formulas are convenient for practical use, as they are expressed in terms of elementary functions and are valid for all values of the external magnetic field.
	
	\item Very good agreement is found between the formulas and numerical solutions of the Stoner–Wohlfarth model.
	
	\item The formulas describe hysteresis loops for cases where anisotropy axes are oriented either randomly or at the same angle relative to the field. 
\end{highlights}
	
\begin{keyword}
		Stoner–Wohlfarth model \sep%
		hysteresis loop \sep%
		magnetic nanoparticles \sep%
		single domain \sep%
		analytical formulas \sep%
		astroid
\end{keyword}

\end{frontmatter}

\section{Introduction}\label{Introduction}
Hysteresis is the central feature of ferromagnetic materials and the property exploited in numerous applications.
Despite the fact that many models of magnetic hysteresis have been proposed \cite{Bertotti_1998,Morrish_2001,Jiles_2002, Jiles_2006,Hodgdon_1988,Atherton_1990,Hauser_1994,DellaTorre_1999,DellaTorre_2004,Nicusor_2019,Moree_2023,Mayergoyz_2003,Liorzou_2000}, there is still a need for an accurate and practically convenient description of hysteresis loops.
The well-known Stoner–Wohlfarth (SW) model, which remains in great demand \cite{Stoner_1948,Tannous_2008}, is one of the most convenient models for the theoretical study of magnetic hysteresis. The SW model has been used to describe the magnetic properties of various ferromagnetic systems, including magnetic nanoparticles \cite{Chuev_2007,Tamion_2012,Jubert_2014,Coisson_2019,Komogortsev_2020,Anand_2020, Delgado-Garcia_2021}, permanent magnets \cite{Moree_2022}, ferromagnetic thin films \cite{Coffey_2003, Slonczewski_2009,Alves_2015,Cunado_2017}, and magnetoresistive random access memories \cite{Ikegawa_2020}. The SW model describes both reversible and irreversible changes in the magnetization process. This model can also be a basis for more complex models and algorithms \cite{Dimitropoulos_2006}. Despite its relative simplicity, the SW model is useful for describing, at least qualitatively, hysteresis in many real magnetic systems.

Calculations in the SW model are usually performed numerically. Typical tasks include (i) determining the equilibrium directions of the magnetic moments and (ii) calculating the average magnetic moment by integrating over different orientations of the anisotropy axes relative to the external field.
In the original work \cite{Stoner_1948}, the hysteresis curve was obtained by using numerical integration and inverse interpolation. 

Importantly, some formulas in the SW model can be obtained analytically. This could help to avoid time-consuming numerical calculations. For example, an exact general analytical solution exists for the problem of determining magnetic states that correspond to extreme values of the particle energy \cite{Wood_2009}. Unfortunately, the awkward form of this solution, which employs complex variables, complicates the analysis of magnetization processes in the SW model.   

There are analytical approaches that approximately model the SW hysteresis loops \cite{Zarkevich_2021,Iglesias_2022,Appino_2023}. These methods are fruitful, but they have some disadvantages. For example, the formulas described in \cite{Iglesias_2022} are applicable only to low magnetic fields. The analytical formulas in \cite{Zarkevich_2021} agree well with numerical results over the entire field range. However, these formulas are obtained through a purely numerical parametrization based on least-squares fitting and do not have a direct physical basis.

Appino \cite{Appino_2023} developed a general statistical formulation for SW ensembles that works for arbitrary distributions of anisotropy constants and orientations. However, for the important special case of randomly oriented particles (uniformly distributed anisotropy axes), this general approach is unnecessarily complex. It relies on probability density functions and requires numerical integration, even though the underlying orientation distribution is simple.

The aim of the present work is to derive a new analytical formula for the ensemble-averaged hysteresis loop of a SW system with uniformly distributed anisotropy axes. The derivation of the formula is based on two approximations:
the low-field approximation, when the deviation of the magnetic moments from
the anisotropy axes is small, and the high-field approximation, when the system eventually approaches saturation. These approximations make it possible to explicitly determine the states (directions) of the magnetic moments, thereby allowing analytical integration to obtain ensemble-averaged hysteresis loops. Both approximations give an accurate result in the limiting cases: in zero field, the moments are aligned with one of the anisotropy axis directions (the choice depends on magnetic history); in very high fields, they are aligned with the external field. The peculiarity of the angular dependence of the switching field in the SW model allows the low-field approximation to be used for intermediate field ranges. The final formula is inevitably piecewise, assembled from individual sections, each most suitable for a given range of the external field. Determining the optimal choice of these intervals is one of the essential parts of the final formula derivation. 

The main advantages of our formula are as follows: a clear physical background, as it is derived from explicitly defined physical approximations; it requires no integration and no statistical sampling; it involves only elementary functions; it is applicable at all values of the magnetic field; it is reasonably accurate and relatively simple.

The article is organized as follows. In Section \ref{SWModel}, we briefly recall the formulation of the SW model. Section \ref{AngleTheta} presents the derivation of approximate analytical formulas for the energetically stable states of the magnetic moments that depend on the field history. Section \ref{Hysteresis(theta_H)} applies these formulas to the derivation of approximate expressions for hysteresis loops in the case of a fixed angle between the anisotropy axis and the external field. Section \ref{AverageMagneticMoment} presents the derivation of the formula for the averaged hysteresis loop. To average the hysteresis loops, we obtained exact inverse formulas for the SW astroid (\ref{app1}). Finally, in Section \ref{Discussions}, we check the validity of our formulas by comparing them with numerical and analytical (for selected cases) solutions of the SW equations.

\section{Physical model of SW particle}
\label{SWModel}

\begin{figure}[htbp]
	\centering
	\includegraphics[height=10cm]
	{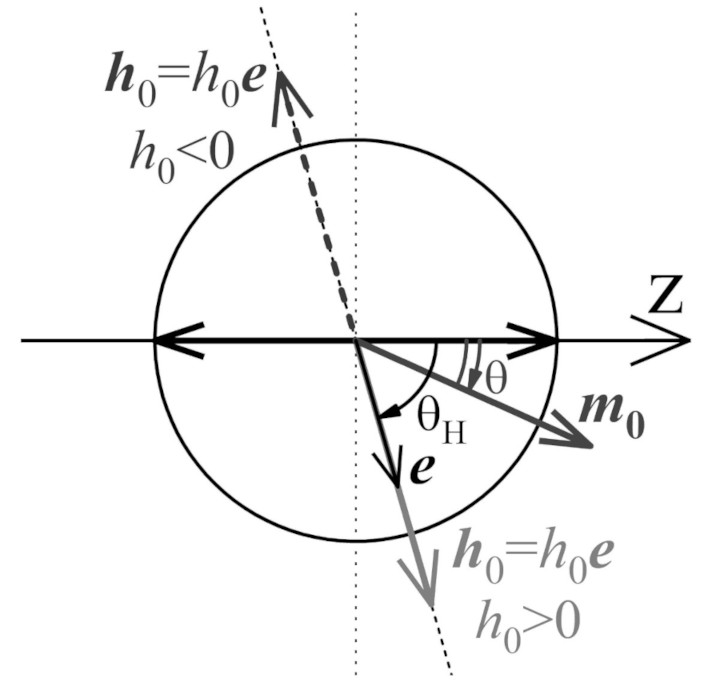}
	\caption{Schematic diagram of a Stoner--Wohlfarth (SW) particle with uniaxial anisotropy along the $z$-axis, placed in a uniform external field $\mathbf{h}_0$. Here, $\mathbf{m}_0$ and $\mathbf{h}_0$ denote the magnetic moment of the particle and applied field, respectively; $\theta_H$ is the angle between the $z$-axis and the positive direction of the external field, defined by the unit vector $\mathbf{e}$; and $\theta$ is the angle between the $z$-axis and the magnetic moment vector $\mathbf{m}_0$.}\label{fig1} 
\end{figure}

We consider a single-domain magnetic particle with uniform magnetization $\mathbf{M}_0$ and uniaxial anisotropy characterized by an effective constant $K_a$. The anisotropy axis is collinear with the $z$-axis (see Fig.~\ref{fig1}).  
Without loss of generality, we assume the particle is a sphere of radius $a$. 
The total magnetic moment is $\mathbf{m}_0 = V_0 \mathbf{M}_0$, where $V_0 = 4\pi a^3/3$ is the volume of the particle.  
The particle is subjected to an external uniform magnetic field $\mathbf{h}_0$, as shown in Fig.~\ref{fig1}. 

Following \cite{Bertotti_1998, Coey_2010, Tannous_2008, Morrish_2001}, the reduced energy of the Stoner–Wohlfarth particle is given by
\begin{equation}\label{SW_Model_Eq1}
	w = \frac{W}{2K_a V_0} = \frac{1}{2}\sin^2\theta - h_0 \cos(\theta - \theta_H),
\end{equation}
where the first and second terms represent the magnetic anisotropy energy and the Zeeman energy, respectively. Here $h_0 = H_0/H_a$ is the reduced external magnetic field, which can take both positive and negative values, and $H_a = 2K_a/M_0$ is the anisotropy field. The angle $\theta \in [0, 2\pi]$ is the polar angle between the direction of the magnetic moment and the positive $z$-axis, measured in the plane containing both the anisotropy axis and the magnetic field (see Fig.~\ref{fig1}). The acute angle $\theta_H \in [0,\pi/2]$ is fixed and represents the angle between the anisotropy axis ($z$-axis) and the direction of the external field $\mathbf{e}$ for $h_0 > 0$, where $\mathbf{h}_0 = h_0 \mathbf{e}$. For $h_0 < 0$, the field is directed opposite to $\mathbf{e}$, which is automatically taken into account by the negative sign of $h_0$ in the Zeeman term. This parametrization allows $\theta_{H}$ to be restricted to the first quadrant without loss of generality, since varying the sign of $h_0$ covers all possible field directions relative to the easy axis.

It should be noted that due to the uniaxial symmetry, the energy (Eq.~\eqref{SW_Model_Eq1}) does not depend on the azimuthal angle $\phi$ of the magnetic moment. Minimization with respect to $\phi$ yields $\phi = \phi_H$, where $ \phi_H$ is the azimuthal angle of the external field. At equilibrium, the magnetic moment lies in the plane defined by the anisotropy axis and the external field. Consequently, the problem reduces to a two-dimensional one described solely by the polar angle $\theta$.

The equilibrium angle $\theta=\theta^{\ast}$ is determined by the energy minimum condition:
\begin{equation}\label{SW_Model_Eq4}
	\left.\frac{\partial{w}}{\partial{\theta}}\right|_{\theta=\theta^{\ast}}=0;\quad \left.\frac{\partial^2{w}}{\partial{\theta^2}}\right|_{\theta=\theta^{\ast}}>0.
\end{equation}

This condition is equivalent to the following algebraic equation:

\begin{equation}\label{SW_Model_Eq5}
	\frac{1}{2}\sin2\theta^{\ast}+h_0\sin(\theta^{\ast}-\theta_{H})=0,
\end{equation}
with the stability condition:

\begin{equation}\label{SW_Model_Eq6}
	\cos2\theta^{\ast}+h_0\cos(\theta^{\ast}-\theta_{H})>0.
\end{equation}

The switching field occurs at the field value for which both the first and second derivatives of the energy $w$ with respect to $\theta$ are zero, i.e., at the inflection point. The switching field $h_{sw}$ is the minimum magnitude of the external magnetic field that causes the magnetic moment vector to jump from a metastable minimum to a stable one. The angular dependence of the switching field (the SW astroid) is described by the explicit formula \cite{Bertotti_1998}:

\begin{equation}\label{SW_Model_Eq7}
	h_{sw}=\frac{1}{\left(\sin^{2/3}\theta_H+\cos^{2/3}\theta_H\right)^{3/2}}.
\end{equation}

Equation~\eqref{SW_Model_Eq5} can be solved numerically. The number of local minima depends on $|h_0|$ and $\theta_H$. For a given $\theta_H$, if $|h_0|<h_{sw}$, Eq.~\eqref{SW_Model_Eq5} has four solutions: two corresponding to local minima and two to local maxima of the energy (Eq.~\eqref{SW_Model_Eq1}). When $|h_0|>h_{sw}$, only two solutions exist: one local minimum and one local maximum. The choice of a specific solution is determined by the magnetic history.

In the next section, we derive approximate formulas for the angle $\theta^{\ast}$ corresponding to the local energy minimum for a given direction and magnitude of the external magnetic field $\mathbf{h}_0$. These formulas are used in Sections 4 and 5 to derive approximate expressions for hysteresis loops. 

\section{Approximate formulas for $\theta^{\ast}$}
\label{AngleTheta}

Based on the low-field and high-field approximations discussed in the Introduction~\ref{Introduction}, we now derive approximate expressions for $\theta^{\ast}$.
Without loss of generality, we consider the ascending branches of the hysteresis loops. For the descending branches, the derivation is analogous. In what follows, we assume $h_{sw} > 0$.

Suppose a field $h_0$ is applied to the particle. The particle has an arbitrary orientation of the uniaxial anisotropy axis and was previously magnetized to saturation by a negative field with respect to the $z$-axis ($h_0 < 0$). Consequently, the magnetic moment initially points in the direction $\theta^{\ast} \in \left[\pi/2, 3\pi/2\right]$. If the applied field satisfies $h_0 < h_{sw}$, switching does not occur, and the particle remains in the local minimum at $\theta^{\ast} \in \left[\pi/2, 3\pi/2\right]$, which is not the absolute energy minimum when $h_0 > 0$. If instead $h_0 > h_{sw}$, the energy of the particle instantly reaches the absolute minimum, with $\theta^{\ast} \in \left[0, \pi/2\right]$.

We consider three cases based on the relationships between the external field, the anisotropy field, and the switching field:
\begin{enumerate}
	\item The value of the external field is less than the switching field but does not exceed the anisotropy field in magnitude ($h_0 < h_{sw}$, with $|h_0| \lesssim 1$); in this case, $h_0$ may be negative or positive;
	\item The value of the external field is greater than or equal to the switching field but still does not exceed the anisotropy field ($h_0 \geq h_{sw}$, with $|h_0| \lesssim 1$); here $h_0$ is necessarily positive because $h_{sw} > 0$; 
	\item The value of the external field is greater than the anisotropy field ($h_0 > 1$) and eventually approaches saturation ($h_0 \gg 1$).
\end{enumerate}

The following derivations are based on physically meaningful assumptions. When the magnitude of the external magnetic field is less than the anisotropy field ($|h_0| \lesssim 1$), the magnetic moment is close to the easy magnetization axis, corresponding to the low-field approximation. Conversely, when the external field is much greater than the anisotropy field ($|h_0| \gg 1$), the magnetic moment is predominantly aligned with the external field, with a small deviation from it, corresponding to the high-field approximation. All three cases (1--3) are considered separately in the following subsections. Cases (1) and (2) are treated within the low-field approximation (Subsections~\ref{BeforeTheSwitchingField} and~\ref{AfterTheSwitchingField}), while case (3) is treated within the high-field approximation (Subsection~\ref{StrongExternalField}).

\subsection{Before switching: $h_0 < h_{sw}$, with $|h_0| \lesssim 1$}
\label{BeforeTheSwitchingField}

If the external field $h_0$ remains below the switching field $h_{sw}$ ($h_0 < h_{sw}$), the magnetic moment stays aligned in the negative direction with respect to the $z$-axis ($\theta^{\ast} \in [\pi/2, 3\pi/2]$). If $-1 < h_0 < 0$, the moment direction remains close to the anisotropy axis ($\theta^{\ast} \to \pi$), which is energetically preferable. If $0 < h_0 < h_{sw}$, the moment points in the negative direction with respect to the $z$-axis in the vicinity of the anisotropy axis. In this case, the external field is too small to overcome the energy barrier and switch the moment to the absolute energy minimum. If $\theta_H$ is close to $\pi/2$, the moment direction can significantly deviate from the anisotropy axis. A quantitative analysis of the deviation is presented in Section~\ref{Discussions}. 

It should be noted that the direction of the moment is negative with respect to the $z$-axis, not relative to the external field direction. For a given $\theta_H$, the projection of the moment onto the $z$-axis may be negative, while its projection onto the external field direction may be positive. For example, just before switching ($h_0 \lesssim h_{sw}$) for $\theta_H$ in the range $(\pi/4, \pi/2]$, the moment has a positive projection onto the direction of the magnetic field. At $\theta_H = \pi/2$, the magnetization reverses continuously without a jump; the switching field is $h_{sw} = 1$, but there is no discontinuous switching. For $\theta_H \in (\pi/4, \pi/2)$ a jump occurs from one positive projection of the magnetic moment onto the field direction to another — specifically, to a projection larger in magnitude. As a result, the coercivity of the averaged hysteresis loop in the SW model is not equal to the smallest switching field value of 0.5 (for $\theta_H  = \pi/4$), but is slightly less: $h_c \approx 0.48$ \cite{Stoner_1948,Coey_2010}.
	
To derive the approximate formula for $\theta^{\ast}$ in the present case, it is convenient to introduce a new variable $\theta^{\prime}=\theta^{\ast}-\pi$, which tends to zero as $\theta^{\ast}\to\pi$. 
Using this new variable, we can rewrite Eq.~\eqref{SW_Model_Eq5} as
\begin{equation}\label{Theta_Eq1}
	\frac{1}{2}\sin(2\theta^{\prime}+2\pi)+h_0\left[\sin(\theta^{\prime}+\pi)\cos\theta_{H}-\cos(\theta^{\prime}+\pi)\sin\theta_{H}\right]=0.
\end{equation}
	
If $\theta^{\prime}$ is sufficiently small, we use the approximations 
$\sin\theta^{\prime}\approx\theta^{\prime}$ and $\cos\theta^{\prime}\approx1$, which allow us to obtain from Eq.~\eqref{Theta_Eq1} the following approximate expression for $\theta^{\prime}$:
	
\begin{equation}\label{Theta_Eq2}
	\theta^{\prime}\approx-\frac{h_0\sin\theta_{H}}{1-h_0\cos\theta_{H}}.
\end{equation}
	
Expressing this result in terms of $\theta^{\ast}$, we obtain the approximate formula for $\theta^{\ast}$ in the case $h_0 < h_{sw}$, with $|h_0| \lesssim 1$:
	
\begin{equation}\label{Theta_Eq4}
	\theta^{\ast}\approx\pi-\frac{h_0\sin\theta_{H}}{1-h_0\cos\theta_{H}}.
\end{equation}

A comparable formula based on similar reasoning was obtained in \cite{OGrady_1980}.

\subsection{After switching: $h_0 \geq h_{sw}$, with $|h_0| \lesssim 1$}
\label{AfterTheSwitchingField}

In the case $h_0 > h_{sw}$ (with $h_0 \lesssim 1$), the magnetic moment jumps from a metastable energy minimum to a stable one. After switching, the moment is aligned in the positive direction with respect to the $z$-axis, near the anisotropy axis. Unlike the previous case, the moment now lies in the absolute energy minimum. For $\theta_H$ close to $\pi/2$, the moment deviates significantly from the anisotropy axis. However, these deviations are generally smaller than before switching. Quantitative estimates are discussed in Section~\ref{Discussions}.

Using Eq.~\eqref{SW_Model_Eq5} and taking into account small-angle approximations $\sin\theta^{\ast}\approx\theta^{\ast}$ and $\cos\theta^{\ast}\approx 1$, we obtain the following expression for $\theta^{\ast}$ in the case $h_{sw}\leq h_0\lesssim 1$:
\begin{equation}\label{Theta_Eq5} 
	\theta^{\ast}\approx \frac{h_0\sin\theta_{H}}{1+h_0\cos\theta_{H}}.
\end{equation}

Importantly, Eqs.~\eqref{Theta_Eq4} and~\eqref{Theta_Eq5} are approximate; their error grows with increasing $h_0$ and $\theta_H$. Fortunately, subsequent calculations show that substituting these approximate expressions into the cosine when calculating the projection onto the external magnetic field, and integrating when obtaining the averaged hysteresis loops, noticeably diminishes the deviations of the final formulas.

\subsection{After switching: $h_0 > 1$, with $h_0 \to \infty$}
\label{StrongExternalField}
    
If the external magnetic field is sufficiently high ($h_0\gg1$), the magnetic moment of the particle becomes aligned with the external field, $\theta^{\ast}\to\theta_{H}$. Therefore, it is convenient to introduce a new variable $\theta^{\prime\prime}=\theta_{H}-\theta^{\ast}\to 0$, which represents a small deviation from the field direction, and rewrite  Eq.~\eqref{SW_Model_Eq5} in terms of $\theta^{\prime\prime}$:
    
\begin{equation}\label{Theta_Eq7}
    \frac{1}{2}\sin2\theta_{H}-h_0\sin\theta^{\prime\prime}=0.
\end{equation}
    
Using the small-angle approximation $\sin\theta^{\prime\prime} \approx \theta^{\prime\prime}$, we obtain the following expressions for $\theta^{\prime\prime}$:
    
\begin{equation}\label{Theta_Eq8}
   \theta^{\prime\prime} \approx \frac{\sin\theta_{H}\cos\theta_{H}}{h_0},
\end{equation}
 and for $\theta^{\ast}$:
\begin{equation}\label{Theta_Eq9}
    \theta^{\ast}\approx\theta_{H}-\frac{\sin\theta_{H}\cos\theta_{H}}{h_0}.
\end{equation}

Using Eq.~\eqref{Theta_Eq7}, we obtain the following approximate expression:
\begin{equation}\label{Theta_Eq10}
	\cos(\theta^{\ast}-\theta_{H})\approx \sqrt{1-\left(\frac{\sin2\theta_{H}}{2h_0}\right)^2},
\end{equation}
which will be employed below to calculate the projection of the magnetic moment onto the field direction.
    
Combining Eqs.~\eqref{Theta_Eq4},~\eqref{Theta_Eq5}, and~\eqref{Theta_Eq9}, we arrive at the final expression for $\theta^{\ast}$:
\begin{align}\label{Theta_Eq11}
    \theta^{\ast}\approx
    \left\{
    \begin{array}{rcl}
    	\displaystyle\pi-\frac{h_0\sin\theta_{H}}{1-h_0\cos\theta_{H}},&  h_0 < h_{sw}; |h_0| \lesssim 1;\\
    	\\
    	\displaystyle\frac{h_0\sin\theta_{H}}{1+h_0\cos\theta_{H}},& h_0 \geq h_{sw}; |h_0| \lesssim 1;\\
    	\\
    	\displaystyle\theta_{H}-\frac{\sin\theta_{H}\cos\theta_{H}}{h_0},& h_0\gg1.
    	\\
    \end{array}
    \right.
\end{align}

The obtained equations will be used in the following sections to derive formulas for non-averaged and averaged hysteresis loops.

\section{Hysteresis loop for a fixed $\theta_H$}\label{Hysteresis(theta_H)}   

In this section, we consider the formulas for the hysteresis loops for a fixed angle $\theta_H$ between the anisotropy axis and the external field. These formulas represent a compromise between precision and simplicity, which we have achieved with our approach.

The projection of the magnetic moment onto the positive direction of the external field can be written as follows:
\begin{equation}\label{Hysteresis(theta_H)Eq1}
	m_{H_0}=m_0\cos(\theta^{\ast}-\theta_{H}).
\end{equation}

Substituting Eqs.~\eqref{Theta_Eq4}, \eqref{Theta_Eq5}, and \eqref{Theta_Eq10} into Eq.~\eqref{Hysteresis(theta_H)Eq1}, we obtain approximate formulas for $m_{H_0}/m_0$ as a function of $h_0$ parameterized by $\theta_H$. Generalizing the reasoning from Section~\ref{AngleTheta}, we present these formulas for both the ascending and descending branches of the hysteresis loop over the following intervals of $h_0$:

\begin{enumerate}
	\item Before switching:
	
	$-1.5 \leq h_0 < h_{sw}$, ascending branch
	\begin{equation}\label{Hysteresis(theta_H)Eq2a}
		\frac{m_{H_0}}{m_0}= - \cos\left(\frac{h_0\sin\theta_H}{1 - h_0\cos\theta_H}+\theta_H\right),
	\end{equation}
	
	$-h_{sw} < h_0 \leq 1.5$, descending branch
	\begin{equation}\label{Hysteresis(theta_H)Eq2b}
		\frac{m_{H_0}}{m_0}= \cos\left(\frac{h_0\sin\theta_H}{1 + h_0\cos\theta_H}-\theta_H\right),
	\end{equation}
	
	\item After switching:
	
	$h_{sw} < h_0 \leq 1.5$, ascending branch
	\begin{equation}\label{Hysteresis(theta_H)Eq3a}
		\frac{m_{H_0}}{m_0}= \cos\left(\frac{h_0\sin\theta_H}{1 + h_0\cos\theta_H}-\theta_H\right),
	\end{equation}
	
	$-1.5 \leq h_0 < -h_{sw}$, descending branch
	\begin{equation}\label{Hysteresis(theta_H)Eq3b}
		\frac{m_{H_0}}{m_0}= -\cos\left(\frac{h_0\sin\theta_H}{1 - h_0\cos\theta_H}+\theta_H\right),
	\end{equation}
	
	\item Approaching saturation:
	
	$|h_0|>1.5$, for both the ascending and descending branches
	\begin{equation}\label{Hysteresis(theta_H)Eq4} 
		\frac{m_{H_0}}{m_0}= \operatorname{sgn}(h_0)\sqrt{1-\left(\frac{\sin 2\theta_H}{2h_0}\right)^{2}},
	\end{equation}
	
\end{enumerate}

The joining of Eqs.~\eqref{Hysteresis(theta_H)Eq2a}--\eqref{Hysteresis(theta_H)Eq3b}, and~\eqref{Hysteresis(theta_H)Eq4} at $h_0=1.5$ is chosen based on the best fit to the numerical solution. According to the assumptions made earlier, Eqs.~\eqref{Hysteresis(theta_H)Eq2a}--\eqref{Hysteresis(theta_H)Eq4} are applicable only in the regions $|h_0| \lesssim 1$, i.e., the low-field approximation, and $|h_0|\gg 1$, i.e., the high-field approximation, respectively. However, numerical analysis shows that Eqs.~\eqref{Hysteresis(theta_H)Eq2a}--\eqref{Hysteresis(theta_H)Eq3b} are reasonably precise up to $h_0 = \pm 1.5$. Moreover, Eqs.~\eqref{Hysteresis(theta_H)Eq2a}--\eqref{Hysteresis(theta_H)Eq3b} and Eq.~\eqref{Hysteresis(theta_H)Eq4} yield very similar values at $h_0 = \pm 1.5$.

Figure~\ref{fig2} shows the dependencies of $m_{H_0}/m_0$ on $h_0$ for different values of $\theta_H$.
The solid and dotted curves were obtained numerically from Eqs.~\eqref{SW_Model_Eq5}--\eqref{SW_Model_Eq7}, \eqref{Hysteresis(theta_H)Eq1} and analytically from Eqs.~\eqref{Hysteresis(theta_H)Eq2a}--\eqref{Hysteresis(theta_H)Eq4}, respectively. The numerical results are in very good agreement with the analytical ones, except for the curve at $\theta_H=\pi/2$.
However, for $\theta_H=\pi/2$, an analytical solution can be obtained separately. To this end, one must solve Eqs.~\eqref{SW_Model_Eq5} and \eqref{SW_Model_Eq6} for $\theta^{\ast}$ and substitute the resulting expression into Eq.~\eqref{Hysteresis(theta_H)Eq1}. For $\theta_H=\pi/2$, the exact solution is:

\begin{equation}\label{Hysteresis(theta_H)Eq5}
	\left.\frac{m_{H_0}}{m_0}\right|_{\theta_H=\pi/2} =
	\begin{cases}
		h_0, & |h_0| < |h_{sw}| = 1; \\
		\operatorname{sgn}(h_0), & |h_0| \geq |h_{sw}| = 1.
	\end{cases}
\end{equation}

Similarly, we obtain an expression for the magnetic moment along the external field direction for $\theta_H = 0$:
\begin{equation}\label{Hysteresis(theta_H)Eq6}
	\left.\frac{m_{H_0}}{m_0}\right|_{\theta_H=0} =
	\begin{cases}
		\mp 1, & |h_0| < |h_{sw}| = 1; \\
		\pm 1, & |h_0| \geq |h_{sw}| = 1,
	\end{cases}
\end{equation}
where the upper and lower signs correspond to the ascending and descending branches, respectively. Importantly, for $\theta_H = 0$, Eqs.~\eqref{Hysteresis(theta_H)Eq2a}--\eqref{Hysteresis(theta_H)Eq4} reduce to the same result as Eq.~\eqref{Hysteresis(theta_H)Eq6}.

Averaging the ascending and descending branches in the first and fourth quadrants of the hysteresis loop yields an expression for the initial magnetization curve. Using Eqs.~\eqref{Hysteresis(theta_H)Eq2a} and~\eqref{Hysteresis(theta_H)Eq2b}, we obtain the following equation for the initial magnetization curve, valid for $h_0 < h_{sw}$:

\begin{equation}\label{Hysteresis(theta_H)Eq8}
	\dfrac{m_{H_0, init}}{m_0}= \sin\left(\dfrac{ h_0\sin\theta_H}{1-h_0^2\cos^2\theta_H}\right)\sin\left(\dfrac{ h_0^{2}\sin\theta_H\cos\theta_H}{1-h_0^2\cos^2\theta_H}+\theta_H\right).
\end{equation}

In the next section, we derive approximate analytical formulas for the averaged hysteresis loop of an ensemble of particles with uniformly distributed anisotropy axes.

\section{Averaged hysteresis loop for randomly oriented anisotropy axes}\label{AverageMagneticMoment}

An ensemble of non-interacting particles with uniformly distributed anisotropy axes is an idealization of a real polycrystalline magnet.
For such an ensemble, the normalized average magnetic moment of the SW particle ensemble can be written as follows \cite{Stoner_1948,Tannous_2008}:
\begin{equation}\label{AMM_Eq1}
	\frac{\langle m_{H_0}\rangle}{m_0}=\int_{0}^{\pi/2}\cos(\theta^{\ast}-\theta_H)\sin\theta_Hd\theta_H,
\end{equation}
where $\theta^{\ast}$ is determined by Eqs.~\eqref{SW_Model_Eq5} and~ \eqref{SW_Model_Eq6}, and depends on $\theta_H$ and $h_0$.

The average magnetic moment in Eq.~\eqref{AMM_Eq1} is usually calculated numerically. Evaluating the integral in Eq.~\eqref{AMM_Eq1} is not straightforward, since $\theta^{\ast}$ is an implicit function of $\theta_H$. However, the formulas for $\theta^{\ast}$ derived in Section~\ref{AngleTheta} make it possible to evaluate this integral analytically and obtain an analytical  formula for the average magnetic moment (Eq.~\eqref{AMM_Eq1}).

In Subsections~\ref{Region_I}--\ref{all branches}, we derive an expression for $\langle m_{H_0} \rangle/m_0$ as a piecewise function of $h_0$. We demonstrate this derivation for the ascending branch of the averaged hysteresis loop. For the descending branch, the derivation is analogous. In our derivation, the total range of the field is divided into three regions I, II, and III,  each of which yields a separate part of the piecewise approximation formula. These regions are chosen to provide an acceptable compromise between simplicity and accuracy of the final result. The formula in regions I and II is based on the low-field approximation, while the formula in region III is based on the high-field approximation. The averaged hysteresis loop with the indicated region designations I--III is shown in Fig.~\ref{fig3}.

\subsection{Region I}\label{Region_I}
    
Let us consider the reduced average magnetic moment in region~I, where $h_0 \in [-0.75, 0.5]$ (see Fig.~\ref{fig3}). In this region, the strict inequality $h_0 < h_{sw}$ holds for all possible orientations of the anisotropy axes, and the magnitude of the magnetic field is insufficient to switch the magnetic moments.

Direct substitution of $\theta^{\ast}$ (Eq.~\eqref{Theta_Eq4}) into Eq.~\eqref{AMM_Eq1} does not allow the integral to be evaluated analytically; therefore, we perform the following transformations. Using the trigonometric identity $\cos(\alpha-\beta)=\cos\alpha\cos\beta+\sin\alpha\sin\beta$, we rewrite Eq.~\eqref{AMM_Eq1} as follows:
\begin{equation}\label{AMM_Eq3}
	\frac{\langle m_{H_0} \rangle}{m_0}=\int_{0}^{\pi/2}(\cos\left.\theta^{\ast}\right|_{h_0<h_{sw}}\cos\theta_H + \sin\left.\theta^{\ast}\right|_{h_0<h_{sw}}\sin\theta_H)\sin\theta_Hd\theta_H.
\end{equation}

Since switching does not occur in this case and all magnetic moments predominantly have a direction with $\theta \approx \pi$ (low-field approximation), it is convenient to use the auxiliary variable $\theta^{\prime} = \theta^{\ast} - \pi$ from Subsection~\ref{BeforeTheSwitchingField}. Using this variable, we rewrite Eq.~\eqref{AMM_Eq3} as follows:

\begin{equation}\label{AMM_Eq4}
	\frac{\langle m_{H_0} \rangle}{m_0}=-\int_{0}^{\pi/2}(\cos\theta^{\prime}\cos\theta_H + \sin\theta^{\prime}\sin\theta_H)\sin\theta_Hd\theta_H.
\end{equation}

Taking into account $\sin\theta^{\prime} \approx \theta^{\prime}$ and $\cos\theta^{\prime} \approx 1$, we rewrite Eq.~\eqref{AMM_Eq4} in the form: 

\begin{equation}\label{AMM_Eq5}
	\frac{\langle m_{H_0} \rangle}{m_0}=-\int_{0}^{\pi/2}(\cos\theta_H + \theta^{\prime}\sin\theta_H)\sin\theta_Hd\theta_H.
\end{equation}

Substituting Eq.~\eqref{Theta_Eq2} into Eq.~\eqref{AMM_Eq5}, we obtain

\begin{equation}\label{AMM_Eq6}
	\frac{\langle m_{H_0} \rangle}{m_0}=\int_{0}^{\pi/2}\frac{h_0 - \cos \theta_H}{1 - h_0 \cos \theta_H}\sin\theta_H d\theta_H.
\end{equation}

Introducing the variable $y = \cos\theta_H$, we can rewrite Eq.~\eqref{AMM_Eq6} in a more compact form:

\begin{equation}\label{AMM_Eq7}
		\frac{\langle m_{H_0} \rangle}{m_0}=\int_{0}^{1}\frac{h_0-y}{1-h_0y}dy.
\end{equation}

Evaluating the integral, we obtain
\begin{equation}\label{AMM_Eq8}
    \frac{\langle m_{H_0} \rangle}{m_0}=\ln(1-h_0)\left(\frac{1}{h_0^2}-1\right)+\frac{1}{h_0}.
\end{equation}
    
Expanding the logarithm in Eq.~\eqref{AMM_Eq8} to third order as $\ln(1 - h_0) \approx - h_0 - h_0^2 / 2 - h_0^3 / 3$, we obtain the following desired expression:
\begin{equation}\label{AMM_Eq9}
     \frac{\langle m_{H_0} \rangle}{m_0}=\dfrac{1}{3}h_0^3+\dfrac{1}{2}h_0^2+\dfrac{2}{3}h_0-\dfrac{1}{2}.
\end{equation}

It is important to note that expanding the logarithm in Eq.~\eqref{AMM_Eq8} to third order is optimal. Its absolute deviation from the numerically evaluated integral in Eq.~\eqref{AMM_Eq1} is less than $0.05$ (at $h_0 = 0.5$). 

For comparison, using Eq.~\eqref{AMM_Eq8} gives an absolute deviation of up to $0.13$ (at $h_0 = 0.5$), while a second-order expansion yields up to $0.12$ (at $h_0 = -0.66$). Furthermore, the third-order expansion provides a lower relative deviation for the coercivity ($h_c = 0.5$), with a relative deviation of about $4\%$ from the numerically calculated value ($h_c \approx 0.48$). Without expansion, the relative deviation is $12\%$, and with a second-order expansion it is $18\%$. Thus, the third-order expansion effectively reduces the error associated with the low-field approximation. Absolute deviations are used for the reduced magnetic moment, as the relative deviation diverges near zero.
    
\subsection{Region II}\label{Region_II}

In contrast to Region~I, the magnitude of the magnetic field in Region~II ($h_0 \in (0.5, 0.75]$) (see Fig.~\ref{fig3}) may be sufficient to switch the magnetic moments. In this field range, the magnetic moments are reversed. Each orientation of the anisotropy axis has its own switching field value, given by Eq.~\eqref{SW_Model_Eq7}. Depending on $\theta_H$, the moments can be oriented either parallel or antiparallel to the magnetic field.

Fig.~\ref{fig4} shows the first quadrant of the SW astroid. According to Eq.~\eqref{SW_Model_Eq7} and Fig.~\ref{fig4}, the smallest switching field $h_{sw}=1/2$ occurs at $\theta_H=\pi/4$, and the largest switching field $h_{sw}=1$ occurs at $\theta_H=0$ and $\theta_H=\pi/2$. The range of integration in Eq.~\eqref{AMM_Eq1} can be divided into three sections ($1$, $2$, and $3$), as shown in Fig.~\ref{fig4}. For fixed $h_0$, the magnitude of the magnetic field is insufficient for switching in the first ($0\le\theta_H\le\theta^I_H(h_{0})$) and third ($\theta^{II}_H(h_{0})\le\theta_H\le\pi/2$) regions, whereas in the second region ($\theta^I_H(h_{0})\le\theta_H\le\theta^{II}_H(h_{0})$), the magnetic field enables switching. It is convenient to represent Eq.~\eqref{AMM_Eq1} as the sum of three integrals $I_1$, $I_2$, $I_3$:
    
\begin{equation}\label{AMM_Eq10}
    \frac{\langle m_{H_0} \rangle}{m_0}=I_1+I_2+I_3,
\end{equation}
where:
\begin{equation}\label{AMM_Eq11}
	I_1=\int_{0}^{\theta_H^{I}(h_0)}\cos(\left.\theta^{\ast}\right|_{h_0<h_{sw}}-\theta_H)\sin\theta_Hd\theta_H;
\end{equation}

\begin{equation}\label{AMM_Eq12}
	I_2=\int_{\theta_H^{I}(h_0)}^{\theta_H^{II}(h_0)}\cos(\left.\theta^{\ast}\right|_{h_0>h_{sw}}-\theta_H)\sin\theta_Hd\theta_H;
\end{equation}

\begin{equation}\label{AMM_Eq13}
	I_3=\int_{\theta_H^{II}(h_0)}^{\pi/2}\cos(\left.\theta^{\ast}\right|_{h_0<h_{sw}}-\theta_H)\sin\theta_Hd\theta_H.
\end{equation}

The transformations of Eqs.~\eqref{AMM_Eq11} ($I_1$) and~\eqref{AMM_Eq13} ($I_3$) are carried out in exactly the same way as in Subsection~\ref{Region_I}. Let us consider the transformations of Eq.~\eqref{AMM_Eq12} ($I_2$) separately. First, we rewrite Eq.~\eqref{AMM_Eq12} using trigonometric transformations:
\begin{equation}\label{AMM_Eq14}
	I_2=\int_{\theta_H^{I}(h_0)}^{\theta_H^{II}(h_0)}(\cos\left.\theta^{\ast}\right|_{h_0>h_{sw}}\cos\theta_H + \sin\left.\theta^{\ast}\right|_{h_0>h_{sw}}\sin\theta_H)\sin\theta_Hd\theta_H.
\end{equation}

Applying the low-field approximation, where $\sin\theta^{\ast} \approx \theta^{\ast}$ and $\cos\theta^{\ast} \approx 1$ as in Subsection~\ref{AfterTheSwitchingField}, we obtain

\begin{equation}\label{AMM_Eq15}
	I_2=\int_{\theta_H^{I}(h_0)}^{\theta_H^{II}(h_0)}(\cos\theta_H + \theta^{\ast}\sin\theta_H)\sin\theta_Hd\theta_H.
\end{equation}

Substituting Eq.~\eqref{Theta_Eq5} into Eq.~\eqref{AMM_Eq15}, we obtain a formula that is ready for integration:

\begin{equation}\label{AMM_Eq16}
	I_2=\int_{\theta_H^{I}(h_0)}^{\theta_H^{II}(h_0)}\frac{h_0 + \cos \theta_H}{1+h_0 \cos \theta_H}\sin\theta_Hd\theta_H.
\end{equation}

Using the substitution $y=\cos\theta_H$, we can write the integrals $I_1$, $I_2$, $I_3$ in a compact form:
	
\begin{equation}\label{AMM_Eq17}
	I_1=\int_{y^{I}(h_0)}^{1}\frac{h_0-y}{1-h_0y}dy;
\end{equation}
	
\begin{equation}\label{AMM_Eq18}
	I_2=\int_{y^{II}(h_0)}^{y^{I}(h_0)}\frac{h_0+y}{1+h_0y}dy;
\end{equation}
	
\begin{equation}\label{AMM_Eq19}
	I_3=\int_{0}^{y^{II}(h_0)}\frac{h_0-y}{1-h_0y}dy.
\end{equation}
where: $y^{I}(h_0)=\cos\left(\theta^I_H(h_0)\right)$, $y^{II}(h_0)=\cos\left(\theta^{II}_H(h_0)\right)$.

Carrying out the integration yields the following expressions:
	
\begin{equation}\label{AMM_Eq20}
	I_1=\left(\frac{1}{h_0^2}-1\right)\ln\left(\frac{1-h_0}{1-h_0\cos\theta^I_H}\right)+\frac{1-\cos\theta^I_H}{h_0};
\end{equation}
	
\begin{equation}\label{AMM_Eq21}
	I_2=\left(\frac{1}{h_0^2}-1\right)\ln\left(\frac{1+h_0\cos\theta^{II}_H}{1+h_0\cos\theta^I_H}\right)+\frac{\cos\theta^I_H-\cos\theta^{II}_H}{h_0};
\end{equation}
	
\begin{equation}\label{AMM_Eq22}
	I_3=\left(\frac{1}{h_0^2}-1\right)\ln\left(1-h_0\cos\theta^{II}_H\right)+\frac{\cos\theta^{II}_H}{h_0}.
\end{equation}

Substituting Eqs.~\eqref{AMM_Eq20}--\eqref{AMM_Eq21} into Eq.~\eqref{AMM_Eq10}, we obtain
\begin{equation}\label{AMM_Eq23}
	 \frac{\langle m_{H_0} \rangle}{m_0}=\left(\frac{1}{h_0^2}-1\right)\ln\left[(1-h_0)\frac{1-h_0^2\cos^2\theta^{II}_H}{1-h_0^2\cos^2\theta^{I}_H}\right]+\frac{1}{h_0}.
\end{equation}

Using the relation $\cos^{2}x=(1+\cos2x)/2$, we can rewrite Eq.~\eqref{AMM_Eq23} in the following form:
\begin{equation}\label{AMM_Eq24}
	 \frac{\langle m_{H_0} \rangle}{m_0}=\left(\frac{1}{h_0^2}-1\right)\ln\left[(1-h_0)\frac{2-h_0^2\left(1+\cos 2\theta^{II}_H\right)}{2-h_0^2\left(1+\cos 2\theta^{I}_H\right)}\right]+\frac{1}{h_0}.
\end{equation}
	
The derivation of equations for angles $\theta_H^{I}$ and $\theta_H^{II}$ as a function of $h_{sw}$ is presented in \ref{app1}. Substituting Eqs.~\eqref{eqA15} and~\eqref{eqA16} into Eq.~\eqref{AMM_Eq24} and taking into account the condition for switching points, $h_0=h_{sw}$, we obtain the expression for the reduced average magnetic moment in Region~II: 	
\begin{eqnarray}\label{AMM_Eq25}
	& \dfrac{\langle m_{H_0} \rangle}{m_0}=\dfrac{1}{h_0}+\left(\dfrac{1}{h_0^2}-1\right) \nonumber
	\\ &\times\ln\left[\left(1-h_0\right)\dfrac{2-h_0^2+\sqrt{h_0^4-\dfrac{4}{27}\left(1-h_0^2\right)^3}}{2-h_0^2-\sqrt{h_0^4-\dfrac{4}{27}(1-h_0^2)^3}}\right].
\end{eqnarray}

\subsection{Region III}\label{Region_III}

In Region~III ($h_0 \in [0.75, \infty)$), following the high-field approximation, Eq.~\eqref{Theta_Eq10} is used to evaluate the integral in Eq.~\eqref{AMM_Eq1}. Substituting Eq.~\eqref{Theta_Eq10} into Eq.~\eqref{AMM_Eq1}, we obtain

\begin{equation}\label{AMM_Eq26}
	\frac{\langle m_{H_0} \rangle}{m_0}=\int_{0}^{\pi/2}\sqrt{1-\left(\frac{\sin2\theta_H}{2h_0}\right)^2}\sin\theta_Hd\theta_H.
\end{equation}

Expanding the square root into a power series to second order and integrating the resulting expression yields the formula for the reduced average magnetic moment in Region~III:

\begin{equation}\label{AMM_Eq27}
		\frac{\langle m_{H_0} \rangle}{m_0} =1-\dfrac{1}{15}\dfrac{1}{h_0^2}-\dfrac{1}{315}\dfrac{1}{h_0^4}.
\end{equation}

Equation~\eqref{AMM_Eq27} is consistent with the classical law of approach to saturation in ferromagnets \cite{Akulov_1930, Holstein_1941, Devi_2021}. Having obtained expressions for regions~I, II, III, we now proceed to the complete solution.

\subsection{General solution for all values of $h_0$}\label{all branches}

Generalizing the reasoning from Sections~\ref{Region_I}--\ref{Region_III} to the case of a descending branch of the hysteresis loop, we represent the final piecewise formula for the reduced magnetic moment by choosing the intervals of $h_0$ as a compromise between simplicity and accuracy.

Region~I:

$h_0 \in [-0.75, 0.5]$, ascending branch:
\begin{equation}\label{AMM_Eq28a}
	\frac{\langle m_{H_0} \rangle}{m_0}
	=
	\frac{1}{3}h_0^3
	+ \frac{1}{2}h_0^2
	+ \frac{2}{3}h_0
	- \frac{1}{2}.
\end{equation}

$h_0 \in [-0.5, 0.75]$, descending branch:
\begin{equation}\label{AMM_Eq28b}
	\frac{\langle m_{H_0} \rangle}{m_0}
	=
	\frac{1}{3}h_0^3
	- \frac{1}{2}h_0^2
	+ \frac{2}{3}h_0
	+ \frac{1}{2}.
\end{equation}

Region~II:

$h_0 \in (0.5, 0.75]$, ascending branch:
\begin{equation}\label{AMM_Eq29a}
	\dfrac{\langle m_{H_0} \rangle}{m_0}
	=
	\dfrac{1}{h_0}
	+
	\left( \frac{1}{h_0^2} - 1 \right)
	\ln
	\left[
	(1 - h_0)
	\dfrac{
		2 - h_0^2 + \sqrt{ h_0^4 - \dfrac{4}{27}(1 - h_0^2)^3 }
	}{
		2 - h_0^2 - \sqrt{ h_0^4 - \dfrac{4}{27}(1 - h_0^2)^3 }
	}
	\right].
\end{equation}

$h_0 \in [-0.75, -0.5)$, descending branch:
\begin{equation}\label{AMM_Eq29b}
	\dfrac{\langle m_{H_0} \rangle}{m_0}
	=
	\dfrac{1}{h_0}
	-
	\left( \dfrac{1}{h_0^2} - 1 \right)
	\ln
	\left[
	(1 + h_0)
	\dfrac{
		2 - h_0^2 + \sqrt{ h_0^4 - \dfrac{4}{27}(1 - h_0^2)^3 }
	}{
		2 - h_0^2 - \sqrt{ h_0^4 - \dfrac{4}{27}(1 - h_0^2)^3 }
	}
	\right].
\end{equation}

Region~III:

$h_0 \in [0.75, \infty)$, ascending branch:
\begin{equation}\label{AMM_Eq30a}
	\frac{\langle m_{H_0} \rangle}{m_0}
	=
	1
	- \frac{1}{15h_0^2}
	- \frac{1}{315h_0^4}.
\end{equation}

$h_0 \in (-\infty, -0.75]$, descending branch:
\begin{equation}\label{AMM_Eq30b}
	\frac{\langle m_{H_0} \rangle}{m_0}
	=
	-1
	+ \frac{1}{15h_0^2}
	+ \frac{1}{315h_0^4}.
\end{equation}

Figure~\ref{fig3} shows the dependencies of $\langle m_{H_0} \rangle / m_0$ on $h_0$. The dots represent the analytical solution in regions~I--III (Eqs.~\eqref{AMM_Eq28a}--\eqref{AMM_Eq30b}), while the solid line shows the numerical solution. One can see that the approximate analytical solution is in good agreement with the numerical one.

By averaging Eqs.~\eqref{AMM_Eq28a}--\eqref{AMM_Eq30b} for the ascending and descending branches in the first and fourth quadrants, we obtain expressions for the initial magnetization curve. For example, the expression for the initial magnetization curve in the range $h_0 \in [0, 0.5]$ is

\begin{equation}\label{AMM_Eq31}
	\frac{\langle m_{H_0} \rangle_{init}}{m_0}=\frac{2}{3}h_0 + \frac{1}{3}h_0^3.
\end{equation}

The full curve (including the region $h_0 > 0.5$) is shown in Fig.~\ref{fig3}, where the analytical expression (dotted curve) becomes more cumbersome and is therefore omitted here. The solid line corresponds to the numerical solution. 

In the next section, we discuss the advantages, shortcomings and accuracy of our formulas by comparing them with numerical and analytical (for selected cases) solutions of the SW equations.

\section{Discussions}\label{Discussions}

In this work, we have presented approximate explicit formulas for hysteresis loops in the SW model. The formulas were obtained for hysteresis loops with a fixed angle $\theta_H$ between the anisotropy axis and the external field (Section~\ref{Hysteresis(theta_H)}), as well as for averaged loops with randomly oriented anisotropy axes (Section~\ref{AverageMagneticMoment}).

A simple physical picture guided our derivation. Specifically, we made the following assumption: when the external field magnitude is less than the anisotropy field (low-field approximation), the magnetic moment lies near the anisotropy axis; when it exceeds the anisotropy field (high-field approximation), the magnetic moment lies near the direction of the external field. The choice of the specific direction along the uniaxial anisotropy axis was determined by the magnetic history and by whether the external field is above or below the switching field.

The proposed assumptions allowed us to obtain approximate formulas for the equilibrium angle $\theta^{\ast}$ of the magnetic moment in three cases (see Section~\ref{AngleTheta}): the external field value is (i) $h_0 < h_{sw}$, with $|h_0| \lesssim 1$ (Eq.~\eqref{Theta_Eq4}); (ii) $h_0 \geq h_{sw}$, with $|h_0| \lesssim 1$ (Eq.~\eqref{Theta_Eq5}); (iii) $h_0 > 1$, with $h_0 \to \infty$ (Eq.~\eqref{Theta_Eq9}).

The magnetization of particles in the SW model changes abruptly at the switching field $h_{0} = h_{sw}$, except for the case $\theta_H=\pi/2$. Therefore, it is natural to use the astroid (Fig.~\ref{fig4}) as a convenient boundary between regions in which different formulas are used to calculate the magnetization. Near  this curve, we employ the low-field approximation (Eqs.~\eqref{Theta_Eq4} and~\eqref{Theta_Eq5}), which assumes a small deviation of the magnetic moment from the anisotropy axis.

One can check how well this approach holds for the solutions to Eq.~\eqref{SW_Model_Eq5} at points on the astroid with $\theta_H$ ranging from $0$ to $\pi/2$.

There are two types of such solutions. The first type describes solutions for which both the first and second derivatives of the free energy vanish. For these solutions, the following conditions hold:

\begin{equation}\label{Discussion_Eq1}
	h_{\parallel}=-\cos^3 \theta^{a},
\end{equation}

\begin{equation}\label{Discussion_Eq2}
	h_{\perp}=\sin^3 \theta^{a},
\end{equation}
where
\begin{equation}\label{Discussion_Eq3}
	h_{\parallel} = h_{sw} \cos \theta_{H},
\end{equation}

\begin{equation}\label{Discussion_Eq4}
	h_{\perp} = h_{sw} \sin \theta_{H},
\end{equation}
and $h_{sw}$ is given by Eq.~\eqref{SW_Model_Eq7}.

From Eqs.~\eqref{Discussion_Eq1} and~\eqref{Discussion_Eq3}, we obtain the following exact solution:

\begin{equation}\label{Discussion_Eq5}
	\theta^{a} = \arccos\left(-\frac{(\cos\theta_H)^{1/3}}{(\sin^{2/3}\theta_H + \cos^{2/3}\theta_H)^{1/2}}\right).
\end{equation}

The angle $\theta^{a}$ determines the direction of the magnetic moment of the particle just before switching. It corresponds to the angular coincidence of two or three of the four energy extrema. For example, at $\theta_H=0$, $\theta^{a} = \pi$ corresponds to the coincidence of two equivalent local energy maxima and one local energy minimum that is not the absolute minimum. At $\theta_H=\pi/2$, $\theta^{a} = \pi/2$ corresponds to the coincidence of two equivalent local minima and one local maximum. For other angles $\theta_H$, there are no equivalent extrema; instead, $\theta^{a}$ marks the merging of two
extrema that disappear upon crossing the astroid from the inside to the outside.

The second type of solution to Eq.~\eqref{SW_Model_Eq5} on the astroid is the angle $\theta^{b}$, which corresponds to one or two of the four energy extrema for which the second derivative with respect to $\theta$ is nonzero. Such an extremum is either a maximum or a minimum, and its angular position changes smoothly upon crossing the astroid line. The value of $\theta^{b}$ as a function of $\theta_H$ can be found numerically from Eq.~\eqref{Discussion_Eq5} using $\theta^{a}(\theta_H)$.

Indeed, with the exception of $\theta_H=\pi/2$ and $\theta_H=0$, the condition that the first derivative of the energy vanishes can generally be written as:

\begin{equation}\label{Discussion_Eq6}
	\frac{h_{\perp}}{\sin \theta} - \frac{h_{\parallel}}{\cos \theta} = 1.
\end{equation}

Using Eqs.~\eqref{Discussion_Eq1} and~\eqref{Discussion_Eq2}, we obtain an implicit equation for $\theta^{b}(\theta_{H})$:

\begin{equation}\label{Discussion_Eq7}
	\frac{\sin^3(\theta^{a}(\theta_H))}{\sin \theta^{b}} + \frac{\cos^3(\theta^{a}(\theta_H))}{\cos \theta^{b}} = 1.
\end{equation}

It is convenient to solve this equation numerically by introducing the coefficient $k$:

\begin{equation}\label{Discussion_Eq8}
	\theta^{b} = k \theta^{a}.
\end{equation}

Then Eq.~\eqref{Discussion_Eq7} takes the form:

\begin{equation}\label{Discussion_Eq9}
	\frac{\sin^3(\theta^{a}(\theta_H))}{\sin (k\theta^{a})} + \frac{\cos^3(\theta^{a}(\theta_H))}{\cos (k\theta^{a})} = 1.
\end{equation}

The dependences $\theta^{a}(\theta_H)$ and $\theta^{b}(\theta_H)$ are shown in Fig.~\ref{fig5}(a). Since the deviation of the moment from the anisotropy axis before switching is given by the angle $\pi - \theta^a$,  its dependence on $\theta_H$ is also shown in Fig.~\ref{fig5}(a). It can be seen that after switching, the low-field approximation introduces a smaller error than before switching.

Since the magnetization is determined by the projection of the magnetic moment
onto the direction of the external field, to understand the source of the error in our formulas, it is advisable to construct the dependence of the difference between these projections on the external field direction. Near the astroid we can do it using $\theta^{\ast}$ obtained in the low-field approximation (Eqs.~\eqref{Theta_Eq4}--\eqref{Theta_Eq5}) and $\theta^{a}$ and $\theta^{b}$ obtained from Eqs.~\eqref{Discussion_Eq5}--\eqref{Discussion_Eq9}:

\begin{equation}\label{Discussion_Eq10}
	\delta_{cos} = \cos(\theta^{\ast,i} - \theta_H) - \cos(\theta^{i} - \theta_H),
\end{equation}
where $i = a,b$; $\theta^{\ast,a}$ is the approximation of the angle $\theta^{\ast}$ at $h_0 = h_{sw}$ just before switching, which corresponds to Eq.~\eqref{Theta_Eq4}, and $\theta^{\ast,b}$ is the approximation of the angle $\theta^{\ast}$ at $h_0 = h_{sw}$ just after switching, which corresponds to Eq.~\eqref{Theta_Eq5}. These dependencies are shown in Fig.~\ref{fig5}(b).

One can see that the low-field approximation after switching yields a very small error for all angles $\theta_H$, except those close to $\pi/2$. Before switching, the deviation (Eq.~\eqref{Discussion_Eq10}) of the low-field approximation near the astroid is noticeably larger.
 
For a given positive value of the external field greater than $1/2$, the contribution to the magnetization comes from both particles that have experienced switching and those that have not. Their relative contribution determines the total error of the average magnetization.

Although the expressions for the equilibrium angle are accurate mainly at low ($h_0 \ll 1$) and high ($h_0 \gg 1$) fields, the resulting hysteresis loop formulas nevertheless retain good accuracy. Let us now examine the errors of the obtained formulas in detail. We use absolute deviations for the reduced magnetic moment, since the relative deviation diverges as the moment approaches zero. To quantify the absolute deviation of the analytical formulas from the numerical solution at a fixed value of $h_0$, we define
\begin{equation}\label{Discussion_Eq11}
	|\delta| =\left| \left( \frac{m_{H_0}}{m_0}\right)_{\mathrm{num}} -\left( \frac{m_{H_0}}{m_0}\right)_{\mathrm{an}} \right|
\end{equation}
where the subscripts "num" and "an" denote the numerically calculated and analytically obtained values of the normalized magnetic moment, respectively.

\begin{figure}[htbp]
	\centering
	\includegraphics[height=10cm]
	{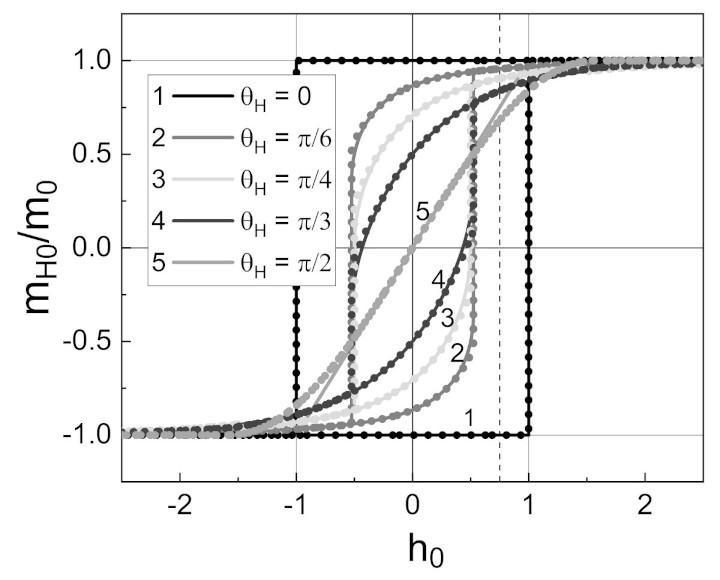}
	\caption{Magnetization curves for a SW particle for fixed angles $\theta_H$. Solid curves represent the results of numerical calculations using Eqs.~\eqref{SW_Model_Eq5}--\eqref{SW_Model_Eq7} and \eqref{Hysteresis(theta_H)Eq1}. Dotted curves represent the analytical solutions obtained from Eqs.~\eqref{Hysteresis(theta_H)Eq2a}--\eqref{Hysteresis(theta_H)Eq4}.}\label{fig2} 
\end{figure}

\begin{figure}[htbp]
	\centering
	\includegraphics[height=10cm]
	{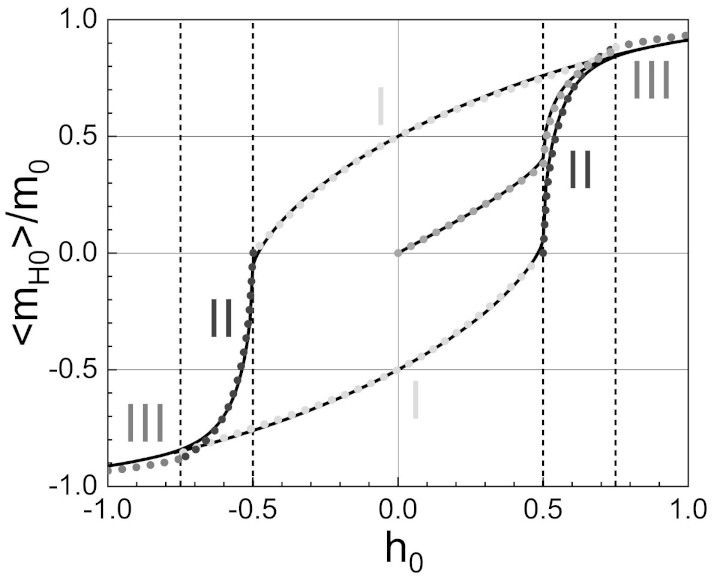}
	\caption{Reduced average magnetic moment $\langle m_{H_0} \rangle / m_0$ as a function of $h_0$. Solid curves show the results of numerical calculations using Eqs.~\eqref{SW_Model_Eq5}--\eqref{SW_Model_Eq7} and \eqref{AMM_Eq1}. Dotted curves are plotted using the following analytical equations: Eqs.~\eqref{AMM_Eq28a} and~\eqref{AMM_Eq28b} in Region I; Eqs.~\eqref{AMM_Eq29a} and~\eqref{AMM_Eq29b} in Region II; Eqs.~\eqref{AMM_Eq30a} and~\eqref{AMM_Eq30b} in Region III.}\label{fig3} 
\end{figure}

Figure~\ref{fig6}(a) shows the dependence of the absolute deviation $|\delta|$ between the analytical formulas (Eqs.~\eqref{Hysteresis(theta_H)Eq2a}--\eqref{Hysteresis(theta_H)Eq4}) for the ascending branch of the non-averaged hysteresis loops and the numerical solution for different angles $\theta_H$. The dotted lines mark the boundaries between the piecewise solutions. The $\theta_H$ values shown in Fig.~\ref{fig6}(a) correspond to the hysteresis loops in Fig.~\ref{fig2}. The absolute deviation $|\delta|$ remains below $0.01$ for all values of $\theta_H$, except in the immediate vicinity of the switching field $h_{sw}$, where it can reach values of up to $0.3$. For angles very close to $\theta_H = \pi/2$, the region of elevated error around $h_{sw}$ broadens (for $\theta_H = \pi/2$, $h_{sw} = 1$), and the absolute deviation $|\delta|$ can reach values of up to $0.16$ (see, for example, the hysteresis loop at $\theta_H = \pi/2$ in Fig.~\ref{fig2}). Fig.~\ref{fig6}(a) demonstrates that the choice of the boundary between the piecewise solutions at $h_0=\pm1.5$ is justified.

Figure~\ref{fig6}(b) shows the field-averaged absolute deviation $\langle|\delta|\rangle$ (left axis) and the maximum absolute deviation $|\delta|_{max}$ (right axis) between the approximate analytical and numerical solutions as functions of $\theta_H/\pi$. The averaged deviation $\langle|\delta|\rangle$ increases non-monotonically with $\theta_H$, consistent with the fact that Eq.~\eqref{Theta_Eq11} is most accurate at smaller angles $\theta_H$. The non-monotonicity arises because the absolute deviation at the boundary between the piecewise formulas at $h_0 =\pm 1.5$ first increases and then decreases (see Fig.~\ref{fig6}(a)). The maximum absolute deviation $|\delta|_{max}$, which occurs near the switching field, also varies non-monotonically. The smallest absolute deviations are observed at small angles $\theta_H$. 

Figure~\ref{fig7} shows the absolute deviation $|\delta|$ between the analytical (Eqs.~\eqref{AMM_Eq28a}--\eqref{AMM_Eq30b}) and numerical solutions for the ascending branch of the averaged hysteresis loop as a function of $h_0$. The dotted lines mark the boundaries between the piecewise solutions. Notably, the most suitable boundary between the piecewise functions corresponds to $h_0= \pm 0.75$. In general, the absolute deviation $|\delta|$ of the obtained analytical formulas does not exceed 0.05, except near the coercivity $h_c = 0.5$, where it increases sharply. This point corresponds to the minimum value of the switching field (at $\theta_H = \pi/4$). Starting from this field value, magnetic moments with other values of $\theta_H$ undergo a cascade switching. The increased absolute deviation near $h_0 = 0.5$ arises because, according to numerical calculations, the magnetization changes sign earlier, at $h_c\approx 0.48$, whereas our analytical formulas yield $h_c = 0.5$.

Even though the accuracy of the approximate formulas for $\theta^{\ast}$ drops in the vicinity of the anisotropy field $h_0 = 1$ ($0.5 \lesssim h_0 \lesssim 1.5$) (see Section~\ref{AngleTheta}), substituting them into the cosine reduces the  deviation of the resulting non-averaged hysteresis loop formulas (Eqs.~\eqref{Hysteresis(theta_H)Eq2a}--\eqref{Hysteresis(theta_H)Eq4}). This is because the bounded range of the cosine $[-1, 1]$ suppresses the growing deviations of $\theta^*$. For the averaged hysteresis loop, the averaging compensates for the errors, and the resulting formulas (Eqs.~\eqref{AMM_Eq28a}--\eqref{AMM_Eq30b}) exhibit reasonable deviations.

Although our formulas are approximate, they take a simple analytical form that is easy to use, unlike the corresponding formulas for SW hysteresis loops presented in the literature \cite{Appino_2023,Zarkevich_2021}.

As an illustration, our formulas allow us to calculate expressions for the initial susceptibility. Since the maximum torque occurs when the external field is perpendicular to the easy-axis orientation of the particle~\cite{Iglesias_2022}, the maximum initial susceptibility of the SW model can be obtained by setting $\theta_H = \pi/2$ in Eq.~\eqref{Hysteresis(theta_H)Eq8}, rewritten in terms of magnetization, and taking the first derivative of the resulting expression with respect to $H_0$, where $H_0 \to 0$. The result coincides with Eq.~(33) in~\cite{Iglesias_2022}:

\begin{equation}\label{Hysteresis(theta_H)Eq9}
	\chi_{\text{max}}=\frac{M_0^2}{2K_a}.
\end{equation}

Our Eq.~\eqref{Hysteresis(theta_H)Eq8} differs from Eq.~(29) in \cite{Iglesias_2022} because we expanded the second sine term in Eq.~\eqref{SW_Model_Eq5} with respect to $\theta$, whereas $\theta$ was taken to be zero in \cite{Iglesias_2022}.

Similarly, for a system of SW particles with randomly oriented anisotropy axis, differentiating Eq.~\eqref{AMM_Eq31} yields an expression for the initial magnetic susceptibility that also coincides with Eq.~(32) in \cite{Iglesias_2022}:

\begin{equation}\label{Hysteresis(theta_H)Eq10}
	\chi_{\text{rand}}=\frac{M_0^2}{3K_a}.
\end{equation}

Thus, despite the local increase in error near the switching field for the hysteresis loops at a fixed $\theta_H$, and near $h_0 \approx 0.5$ for the averaged hysteresis loop, our analytical and rather simple formulas can be useful.

\section{Conclusions}

In this work, we have derived approximate explicit formulas for the hysteresis loops in the Stoner–Wohlfarth model. Both cases were considered: a fixed angle $\theta_H$ between the anisotropy axis and the external field, and an average over randomly oriented anisotropy axes.

The proposed formulas are piecewise functions, with different parts defined over several intervals of the magnetic field. The boundaries between the intervals were chosen to minimize the formula inaccuracy. The derivation is based on a clear physical assumption: the particle magnetic moment lies in the vicinity of the easy axis at low fields and near the external field direction at high fields. Despite the approximate nature of this assumption, the resulting expressions for the magnetization show good agreement with numerical solutions.

For hysteresis loops for a fixed $\theta_{H}$, the absolute deviation remains below 0.05 for most field values, except near the switching field where it can locally reach up to $0.3$. For the averaged hysteresis loop, the maximal deviation of our formula from the numerically calculated magnetization occurs at $h_0 \approx 0.5$, where the sharp change in magnetization slope occurs.

Unlike existing analytical approaches, our formulas are based on a simple physical picture, involve only elementary functions, and are valid over the whole field range. Their analytical form makes them a convenient alternative to numerical calculations.

\section*{CRediT authorship contribution statement}

\textbf{Vladimir P. Savin}: Writing – original draft, Visualization, Validation, Methodology, Investigation, Formal analysis, Conceptualization.
\textbf{Yury A. Koksharov}: Writing – review $\&$ editing, Validation, Supervision, Methodology, Conceptualization.

\section*{Funding}

This research did not receive any specific grant from funding agencies in the public, commercial, or not-for-profit sectors.

\section*{Declaration of competing interest}

The authors declare that they have no known competing financial interests or personal relationships that could have appeared to influence the work reported in this paper.

\section*{Acknowledgment}

The study was conducted under the state assignment of Lomonosov Moscow State University.

\newpage

\newpage

\begin{figure}[htbp]
	\centering
	\includegraphics[height=10cm]
	{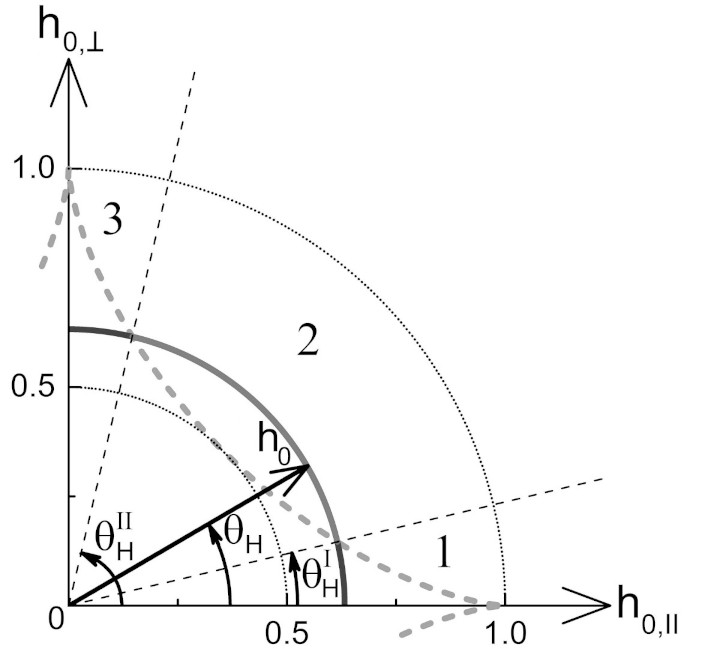}
	\caption{The dependence of $h_{0,\perp}$ on $h_{0,\parallel}$. This scheme illustrates the partition of the integration interval. The thick dotted lines represent the astroid; circular arcs correspond to areas with a constant applied field magnitude ($h_0 = |\mathbf{h}_0|=\text{const}$). The field direction is specified by the angle $\theta_H$. The astroid intersects the circular arc for $0.5 < h_0 \le 1.0$ at angles $\theta^{I}_H$ and $\theta^{II}_H$, which serve as the integration limits in Eqs.~\eqref{AMM_Eq11}--\eqref{AMM_Eq13} and divide the circular arc into three parts (labeled $1$, $2$, and $3$).}\label{fig4} 
\end{figure}

\begin{figure}[htbp]
	\centering
	\includegraphics[height=15cm]
	{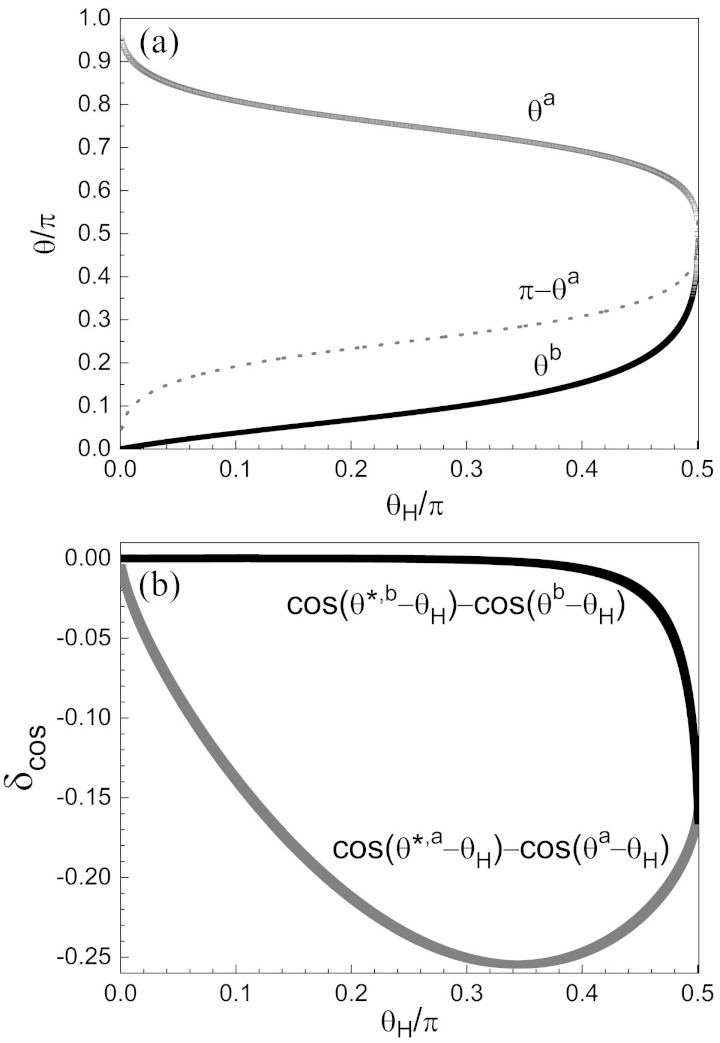}
	\caption{(a) The angles $\theta^{a}$ and $\theta^{b}$ versus $\theta_H$. The dotted line corresponds to $\pi - \theta^{a}$, which represents the deviation of the magnetic moment from the anisotropy axis just before switching. The angle $\theta^{a}$ was calculated analytically using Eq.~\eqref{Discussion_Eq5}. The numerical solution for $\theta^{b}$ was obtained using Eqs.~\eqref{Discussion_Eq8} and~\eqref{Discussion_Eq9}. (b) The deviations $\delta_{cos}$ (defined by Eq.~\eqref{Discussion_Eq10}) versus $\theta_H$.}\label{fig5} 
\end{figure}

\begin{figure}[htbp]
	\centering
	\includegraphics[height=15cm]
	{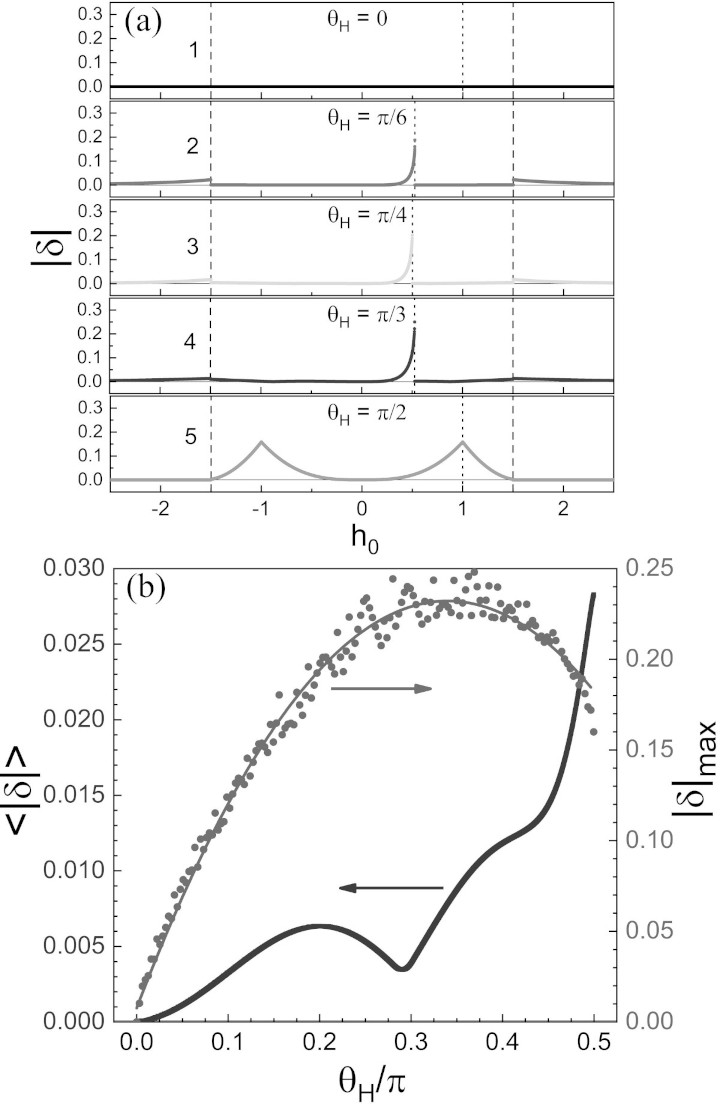}
	\caption{(a) Absolute deviation $|\delta|$ between the analytical solution (Eqs.~\eqref{Hysteresis(theta_H)Eq2a}--\eqref{Hysteresis(theta_H)Eq4}) and the numerical solution for the ascending branches of hysteresis loops versus  $h_0$ for fixed $\theta_H$ (numbers $1$--$5$ correspond to the magnetization curves in Fig.~\ref{fig2}). Dashed lines mark the joining areas of the piecewise analytical formulas; dotted lines show the magnitude of the switching field. (b) Averaged over the range $h_0 \in [-2.5, 2.5]$, the mean absolute deviation $\langle|\delta|\rangle$ (left axis) and the maximum absolute deviation $|\delta|_{max}$ (right axis) between the  analytical formulas (Eqs.~\eqref{Hysteresis(theta_H)Eq2a}--\eqref{Hysteresis(theta_H)Eq4}) and the numerical solution for the ascending branches of hysteresis loops versus $\theta_H/\pi$. The thin solid line is a guide for the eye. The numerical solution was performed using Eqs.~\eqref{SW_Model_Eq5}--\eqref{SW_Model_Eq7} and \eqref{Hysteresis(theta_H)Eq1}.}\label{fig6} 
\end{figure}

\begin{figure}[htbp]
	\centering
	\includegraphics[height=10cm]
	{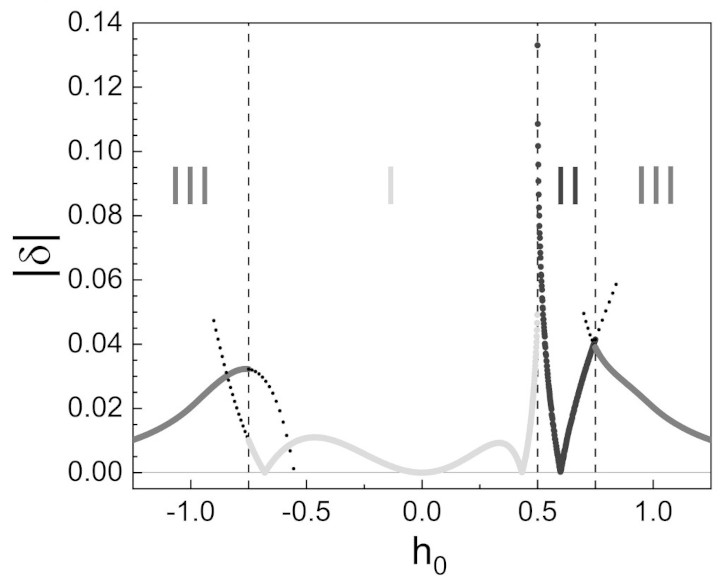}
	\caption{Absolute deviation $|\delta|$ between the approximate analytical solution and the numerical solution for the ascending branch of the averaged hysteresis loop (see Fig.~\ref{fig3}) versus the reduced external field $h_0$. The analytical solution is given by Eqs.~\eqref{AMM_Eq28a}--\eqref{AMM_Eq30b}. The numerical solution was performed using Eqs.~\eqref{SW_Model_Eq5}--\eqref{SW_Model_Eq7} and \eqref{AMM_Eq1}. Dashed lines mark the joining areas of the piecewise analytical formula in the regions I--III (see Fig.~\ref{fig3}).}\label{fig7} 
\end{figure}

\newpage
	
\appendix
\section{Inverse astroid equation}\label{app1}

In this appendix, we derive the inverse formulas for the switching field, giving the $\theta_H$ angle as a function of the field $h_{sw}$ (the inverse of the usual $h_{sw}$ versus $\theta_H$ dependence). These formulas are used to derive the averaged hysteresis in Subsection \ref{Region_II}. 

Let us recall the usual formula for the SW astroid (Eq.~\eqref{SW_Model_Eq7}):
\begin{equation}\label{eqA1}
	h_{sw}=\frac{1}{\left(\sin^{2/3}\theta_H+\cos^{2/3}\theta_H\right)^{3/2}}.
\end{equation}

Equation~\eqref{eqA1} can be written in the form:
\begin{equation}\label{eqA2}
	\left(h_{sw}\cos\theta_H\right)^{2/3}+\left(h_{sw}\sin\theta_H\right)^{2/3}=1.
\end{equation}

It is convenient to introduce the following variables:
\begin{equation}\label{eqA3}
	x=h_{sw}\cos\theta_H,
\end{equation}
\begin{equation}\label{eqA4}
	y=h_{sw}\sin\theta_H.
\end{equation}

Let us rewrite Eq.~\eqref{eqA2} in terms of new the variables $x$ and $y$:

\begin{equation}\label{eqA5}
	x^{2/3}+y^{2/3}=1.
\end{equation}

Raising the both sides of Eq.~\eqref{eqA5} to the third power, we obtain:

\begin{equation}\label{eqA6}
	x^2+3x^{4/3}y^{2/3}+3x^{2/3}y^{4/3}+y^2=1.
\end{equation}

Taking the common factor $3x^{2/3}y^{2/3}$ of the second and third terms out of the parentheses, we get:

\begin{equation}\label{eqA7}
	x^2+y^2+3x^{2/3}y^{2/3}\left(x^{2/3}+y^{2/3}\right)=1.
\end{equation}

Since the expression in parentheses equals one according to Eq.~\eqref{eqA4}, we rewrite Eq.~\eqref{eqA7} as

\begin{equation}\label{eqA8}
	x^2+y^2-1=-3x^{2/3}y^{2/3}.
\end{equation}

Raising Eq.~\eqref{eqA8} to the third power, we obtain the equation of the astroid in the form:

\begin{equation}\label{eqA9}
	(x^2+y^2-1)^3+27x^2y^2=0.
\end{equation}

Substituting the variables $x$ (Eq.~\eqref{eqA3}) and $y$ (Eq.~\eqref{eqA4}) back  into Eq.~\eqref{eqA9} yields:

\begin{equation}\label{eqA10}
	(h_{sw}^2-1)^3+27h_{sw}^4\cos^2\theta_H\sin^2\theta_H=0.
\end{equation}

Moving the terms with the angle $\theta_H$ to the left and the terms with the switching field $h_{sw}$ to the right, we obtain the following equation:

\begin{equation}\label{eqA11}
	\sin^{2}2\theta_H=\frac{4}{27}\frac{(1-h_{sw}^2)^3}{h_{sw}^4}.
\end{equation}

Since we are interested in the interval $0\le\theta_H\le\pi/2$ where $sin^{2}2\theta_H\ge0$, we take the root of Eq.~\eqref{eqA11}:

\begin{equation}\label{eqA12}
	\sin2\theta_H=2\sqrt{\frac{(1-h_{sw}^2)^3}{27h_{sw}^4}}.
\end{equation}

Equation~\eqref{eqA12} yields two roots in the interval $0\le\theta_H\le\pi/2$:

\begin{equation}\label{eqA13}
	\theta_H^{I}=\frac{1}{2}\arcsin\left(2\sqrt{\frac{(1-h_{sw}^2)^3}{27h_{sw}^4}}\right),
\end{equation}

\begin{equation}\label{eqA14}
	\theta_H^{II}=\frac{\pi}{2}-\frac{1}{2}\arcsin\left(2\sqrt{\frac{(1-h_{sw}^2)^3}{27h_{sw}^4}}\right).
\end{equation}

The obtained angles $\theta_H^{I}$ (Eq.~\eqref{eqA13}) and $\theta_H^{II}$ (Eq.~\eqref{eqA14}) delimit the integration range in Eqs.~\eqref{AMM_Eq8}--\eqref{AMM_Eq10} (see Fig.~\eqref{fig4}). From Eqs.~\eqref{eqA13} and~\eqref{eqA14}, we also obtain useful expressions for the cosines of these angles:

\begin{equation}\label{eqA15}
	\cos2\theta^{I}_H=\sqrt{1-\frac{4}{27}\frac{\left(1-h_{sw}^2\right)^3}{h_{sw}^4}},
\end{equation}

\begin{equation}\label{eqA16}
	\cos2\theta^{II}_H=-\sqrt{1-\frac{4}{27}\frac{\left(1-h_{sw}^2\right)^3}{h_{sw}^4}}.
\end{equation}

Equations~\eqref{eqA15} and~\eqref{eqA16} are used for substitution into Eq.~\eqref{AMM_Eq18} in Subsection \ref{Region_II}.

\newpage
	
\bibliographystyle{elsarticle-num}

\end{document}